\documentclass[11pt, letterpaper]{article}

\usepackage{graphicx}
\usepackage{amsmath}
\usepackage{amssymb}
\usepackage{setspace}
\usepackage{epstopdf}
\usepackage{color}
\usepackage[letterpaper,left=1in,right=1in,top=1in,bottom=1in]{geometry}
\usepackage{multirow}
\usepackage{longtable}
\usepackage{rotating}
\usepackage{setspace}
\usepackage{geometry}
\usepackage{subcaption}
\usepackage{amsmath}
\usepackage{algpseudocode}
\usepackage{algorithm}
\usepackage{algorithm,algpseudocode}
\usepackage[definethebibliography]{easybib}
\usepackage{algorithm,algcompatible,amsmath}
\algnewcommand\Initialization{\item[\textbf{Initialization:}]}%

\usepackage{pgfplots}
\pgfplotsset{compat=1.9}
\usetikzlibrary{plotmarks}
\usepackage{float}

\newcommand{\m}{\mathbb}
\title{\Large \bf Subsystem decomposition and state estimation of nonlinear processes with implicit time-scale multiplicity}

\author{
\centerline{\normalsize Sarupa Debnath$^{a}$, Soumya Ranjan Sahoo$^{a}$, Benjamin Decardi-Nelson$^{a}$, Jinfeng Liu$^{a,}$\thanks{Corresponding author: J. Liu. Tel: +1-780-492-1317. Fax: +1-780-492-2881. Email: jinfeng@ualberta.ca.}}
\vspace{3mm}\\
\centerline{\small $^{a}$Department of Chemical \& Materials Engineering, University of Alberta,}\\
\centerline{\small Edmonton, AB T6G 1H9, Canada}}
\allowdisplaybreaks

\begin{document}
\date{}
\maketitle
\begin{abstract}
In this work, we propose a subsystem decomposition approach and a distributed estimation scheme for a class of implicit two-time-scale nonlinear systems. Taking the advantage of the two-time-scale separation, these processes are decomposed into smaller subsystems such as fast subsystem and slow subsystem. In the proposed method, an approach, composite solution, combines the approximate solutions obtained from both fast and slow subsystems. Based on the fast and slow subsystems, a distributed state estimation scheme is proposed to handle the implicit time-scale multiplicity. In the proposed design, an extended Kalman filter (EKF) is designed for the fast subsystem and a moving horizon estimator (MHE) is designed for the slow subsystem. There is a communication between the estimators: the slow subsystem is only required to send information to the fast subsystem one-directionally. The fast subsystem estimator does not send out any information. The estimators use different sampling of the state measurements, i.e., fast sampling of the fast state variables is considered in the fast EKF and slow sampling of the slow state variables is considered in the slow MHE. Extensive simulations based on a chemical process are performed to illustrate the effectiveness and applicability of the proposed subsystem decomposition and composite state estimation architecture.
\end{abstract}


\section*{Introduction}
Chemical plants are large-scale and complex process systems. These processes are inherently nonlinear and generally associate with a common feature, time-scale multiplicity. It usually arises due to the strong coupling of the physical and chemical phenomena occurring at disparate time-scales \cite{kumar_nonlinear_2002}. 
Typical examples of the multiple-time-scale processes are a process with large recycle \cite{kumar_nonlinear_2002}, multiple reactions \cite{kumar_singular_1998}, and reactor/separator system \cite{yin_distributed_2017}, where the time-scale multiplicity occurs due to the presence of distinctly different time constants,  multiple fast and slow reactions, and significantly different residence times, respectively. A direct application of standard control or estimation methods without considering time-scale multiplicity may lead to ill-conditioning or even the loss of closed-loop stability \cite{kokotovic_singular_1986}. To deal with such systems, singular perturbation theory provides a natural framework for modeling, stability analysis, model reduction, and controller design for nonlinear two-time-scale processes \cite{kokotovic_singular_1986}. The main requirement for the application of this method is that the process should be modeled in the standard singularly perturbed form, where the separation of fast and slow variables is explicit. The design of fully centralized nonlinear controllers on the basis of the entire process system is impractical in terms of computational burden, high complexity of its dynamic models, and sensitivity to modeling errors and measurement noise. These considerations generate vigorous interest in decentralised (composite fast and slow control) \cite{chen_composite_2012} and distributed control \cite{chen_model_2011} strategies using proportional (P) control \cite{kumar_nonlinear_2002,kumar_nonlinear_2003} or Model predictive control (MPC) \cite{chen_composite_2012,chen_model_2011}. However, there is little attention paid to decentralized or distributed state estimation which is equally important and is closely related to distributed control. 

In \cite{zhang_distributed_2013}, an observer-enhanced distributed moving horizon estimation (DMHE) algorithm was developed for a nonlinear system composed of several subsystems. A common benchmark simulation model, wastewater treatment plant (WWTP) is decomposed into reduced-order subsystems and an iterative DMHE scheme is implemented \cite{yin_subsystem_2018}. It shows iterative DMHE can provide more balanced performance within a much lower computational cost compared to centralized MHE. Recently, two alternative nonlinear observer design approaches, one full-order, and one reduced-order are designed for a two-time-scale system. The full-order observer is designed based on the linearized original model around its stable steady state, whereas, the reduced-order observer is derived based on a lower-dimensional model to reconstruct the slow states which are used to calculate an invariant manifold for the fast state estimation \cite{duan_nonlinear_2020}. Within the singular perturbation framework, a nonlinear system was decomposed into a fast system and several slow subsystems, and DMHE is applied in \cite{yin_distributed_2017}. One directional communication from the slow subsystem MHEs to the fast subsystem MHE is established and also sufficient conditions on the convergence of the estimation error of the DMHE are derived.

It is mentioned earlier that the singular perturbation theory is the standard tool for the analysis of systems with the explicit time-scale where the small parameter $\varepsilon$ directly separates the fast and slow variables. On the contrary, there is a wide range of applications where slow and fast dynamics cannot be associated with distinct process variables. This implicit time-scale multiplicity cannot be expressed in the standard singularly perturbed form where the separation of fast and slow modes is explicit. Also, due to a lack of process knowledge of the fast and slow variables, there is no way to model directly such processes into the standard singularly perturbed form \cite{kumar_singular_1998}.

In this line of research, coordinate change of the two-time-scale system is widely considered to transform into a standard singular perturbation form. In \cite{georgakis_quasi-modal_1977}, the coordinate change construction for linear two-time-scale systems is addressed using modal analysis. A past work on nonlinear ODEs with a small parameter $\varepsilon$ focused on geometric properties to derive a coordinate-free characterization of the time-scale multiplicity \cite{fenichel_geometric_1979} which was subsequently used to find necessary and sufficient geometric conditions for the existence of an $\varepsilon$-independent change of coordinates that provides a standard singularly perturbed representation \cite{marino_geometric_1988}. In \cite{krishnan_connection_1994}, an $\varepsilon$-dependent coordinate change for a class of nonlinear two-time-scale systems was also proposed. A set of results is developed to obtain a standard singularly perturbed
representation from the original two-time-scale process using both $\varepsilon$-independent and $\varepsilon$-dependent coordinate change \cite{kumar_singular_1998}. Further, an input/output linearizing controller was designed on the basis of the slow subsystem. However, for large-scale practical systems, coordinate change is not recommended because it is not unique for any system and also it increases complexities in the modeling. Therefore, a model reduction methodology is considered for deriving nonlinear low-order models of the fast-slow dynamics and a two-tiered controller framework is designed to establish well-coordinated controllers \cite{kumar_nonlinear_2002, kumar_nonlinear_2003, baldea_dynamics_2006}. In \cite{naidu_singular_2001}, a survey of the applications of the theory and techniques of singular perturbations and time-scales is carried out (SPaTS) in guidance and control of aerospace systems. The degenerate problem or unperturbed problem is expanded into reduced (outer) and boundary-layer (inner) models order and systematically solved considering boundary layer correction terms for the standard singular perturbed problem.

Clearly, there are still many open issues in the development of the state estimation for the system with the implicit time-scale multiplicity which is a common occurrence in chemical processes. It is also worth mentioning that current results mainly focused on the state estimation algorithms based on the explicit time-scale form. Therefore, it is necessary to propose a way to model and estimate an implicit time-scale multiplicity. We propose a general estimation framework for a type of nonlinear implicit time-scale multiplicity. To the best of our knowledge, there is no systematic approach that can be used simultaneously to model and estimate for such systems.

Motivated by the above considerations, in the work, we design a systematic subsystem decomposition approach for distributed state estimation of a class of nonlinear implicit time-scale systems with bounded output measurement noise and process disturbances. We borrow the idea of the method of matched asymptotic expansions that has been successfully used in subsystem configuration for distributed state estimation. The system is first decomposed into a reduced-order fast subsystem and a reduced-order slow subsystem considering different limiting conditions on $\varepsilon$. The approximate solutions of the fast and slow subsystems are combined to calculate the composite solution, the states of the actual system are obtained, which approximates the dynamics of the original nonlinear system. Further, a fast EKF is designed for the fast system and a slow MHE is designed for the slow subsystem. The fast EKF and slow MHE form a distributed scheme. It is found that the slow MHE is entirely decoupled from the fast EKF which is a significant difference from the control of two-time-scale systems. The decoupling ensures that only unidirectional information transmission from the slow MHE to the fast MHE is needed and the fast EKF does not send out any information. Also, we make comparisons between the proposed approach and a decentralized scheme/a centralized MHE scheme. The contributions of this work mainly include:
\begin{enumerate}
\item A systematic procedure based on the inner and outer solution to decompose the entire process into two subsystems such as fast and slow subsystems for nonlinear implicit two-time-scale systems with large parameters of the form $\frac{1}{\varepsilon}$.
\item An introduction of overlap solution from the idea of matching to derive a composite solution.
\item A methodology of distributed estimation scheme that is suitable for estimation of the fast and slow subsystem states for the reconstruction of actual states.
\end{enumerate}
To demonstrate the usefulness of the proposed method, a typical chemical process is considered in the simulations and the simulation results demonstrate the effectiveness of the proposed approach.

\section*{Preliminaries}
\subsection*{Notation}

The operator $L_fh$ represents the Lie derivative of function $h$ with respect to function $f$, calculated following $L_fh(x)=\frac{\partial h}{\partial x}f(x)$. $L_f^rh$ represents the $r^{th}$ order Lie derivative of function $f$, denoted by $L_f^rh(x)=L_fL_f^{r-1}h(x)$. Subscript $f$ and $s$ denote fast and slow subsystems respectively unless mentioned. The subscript $ss$ denotes a variable associated with the steady-state. A matrix is full row rank when each of the rows of the matrix are linearly independent and is full column rank when each of the columns of the matrix are linearly independent.

\subsection*{System description}

In this work, we consider a class of nonlinear system that can be described in the following singularly perturbed form \cite{baldea_dynamics_2012}:
\begin{subequations}
\label{eq:generalsin}
\begin{align}
    \dot{x}(t)&=f(x(t))+ g(x(t))u(t)+\frac{1}{\varepsilon}b(x(t))k(x(t))+w(t) \label{eq:1a} \\
    y(t)&=h(x(t))+v(t) \label{eq:1b} \hspace{2mm}
\end{align}
\end{subequations}

where $x \in X \subset R^{n_x}$ is the vector of state variables of independent time variable $t$, $u \in R^{n_u}$ is the vector of the manipulated inputs, $y \in R^{n_y}$ is the vector containing all the measured outputs, $w \in R^{n_x}$ denotes system disturbances and $v \in R^{n_y}$ is measurement noise, the initial condition is $x(0)=x_0$ and $\varepsilon$ is a small parameter, such that $0<\varepsilon<<1$. $f(x)$ and $k(x)$ are analytic vector fields of dimensions $n_x$ and $p_x$ $(p_x<n_x)$, $g(x)$ and $b(x)$ are analytic matrices of dimensions $(n_x \times n_u)$, $(n_x \times p_x)$, respectively. In Eq. (\ref{eq:generalsin}), the term $\frac{1}{\varepsilon}b(x(t))k(x(t))$ corresponds to the fast dynamics of the system \cite{baldea_dynamics_2012}. We consider that the matrix $b(x)$ and the Jacobian $\frac {\partial k(x)}{\partial x}$ have full column and row rank, respectively. Though the condition on the rank of $b(x)$ is not restrictive. Further, it is assumed that output $y(t)$ is continuously measured \cite{kumar_singular_1998}.

Equation (\ref{eq:generalsin}) describes a class of systems where the separation of fast and slow dynamics is not explicit. There is a wide variety of processes that
exhibit time-scale multiplicity. This phenomenon is generally induced by different thermal properties, mass transfer rates, and chemical kinetics of the reaction \cite{baldea_dynamics_2012}. These systems are mainly characterized by the presence of a small parameter $\varepsilon$ in the explicit mathematical models. The reciprocal of such a small parameter gives a very large term that is responsible for the existence of disparate time-scale features. Typical examples of such processes include those with multiple fast and slow reactions or fast heat/mass transfer rates. These processes typically exhibit two distinctly different magnitudes of gains in different input directions and different time constants: a large and dominant time constant and a small-time constant associated with slow and fast dynamics, respectively. The small parameter $\varepsilon$ differentiates the nature of the fast and slow dynamics, and both the dynamics control the speed of the system, as their names suggest. The overall system response is affected by either fast or slow dynamics or both.  \cite{kumar_nonlinear_2003}.

In the remainder of this section, we will introduce how to derive the fast and slow dynamics of system (\ref{eq:generalsin}) using matched asymptotic expansions method. Further, a composite solution is introduced to retrieve the system's complete dynamics through a correction term. 

\subsection*{Two-time-scales decomposition}

System (\ref{eq:generalsin}) describes a class of two-time-scale systems with implicit time-scale separation where each state can have both fast and slow dynamics. Solving such a system is quite different from solving a system with explicit time-scale separation \cite{yin_distributed_2017}. Matched asymptotic expansions is used to decompose and solve the system. It involves finding different approximate solutions or asymptotic expansions, valid for a particular time-scale, and then combining these different solutions to give a single approximate solution valid for the original system. Due to the brisk nature of the fast dynamics, we need a stretched time-scale to capture the dynamics. Conversely, the slow dynamics are sluggish in nature, so a squeezed time-scale is preferably apt for it. 

Specially, in asymptotic expansions, the limiting solutions for fast or slow dynamics are obtained. An outer (reduced layer) approximation is obtained for the slow dynamics and inner (boundary-layer) approximation for the fast dynamics is obtained \cite{kokotovic_singular_1986}. While combining the inner and outer approximations for the actual dynamics, a correction term needs to be subtracted to avoid considering their overlap value twice. In the following, we consider the nominal deterministic system to illustrate how to decompose the fast and slow dynamics. 

\subsubsection*{Decomposition of fast dynamics}

In the inner approximation, a fast (stretched) time $\tau$ is defined as $\tau=\frac{t}{\varepsilon}$. Multiplying both sides of Eq. (\ref{eq:1a}) by $\varepsilon$, converting into $\tau$ time-scale, and considering the limiting case $\varepsilon \rightarrow 0$, we obtain the fast dynamics of system (\ref{eq:generalsin}):
\begin{equation}
   \frac{dx(\tau \varepsilon)}{d\tau} =b(x(\tau \varepsilon))k(x(\tau \varepsilon))
    \label{eq:4.3}
\end{equation}
Denoting $x(\tau \varepsilon)$ as $x_f(\tau)$, the above equation transforms into,
\begin{equation}
   \frac{dx_f}{d\tau}=b(x_f)k(x_f):=f_f(x_f)
    \label{eq:1.4}
\end{equation}
The above system as shown as (\ref{eq:1.4}) approximates the fast dynamics of the original system (\ref{eq:generalsin}). $f_f(x_f)$ is the analytic vector field of the dimension of $n_{x_f}$. If the steady-state of the fast dynamics is $x_{fss}$, it satisfies the following condition:
\begin{align}
     k(x_{fss})&=0, \hspace{2mm} 
     \label{eq:4.40}
\end{align}
We consider systems of the form of Eq. (\ref{eq:generalsin}) for which the matrix $b(x)$ and the Jacobian $\frac {\partial k(x)}{\partial x}$ have full column and row rank, respectively. The condition of full column rank of $b(x)$ ensures that it cannot be zero. However, the condition on the rank of Jacobian $\frac {\partial k(x)}{\partial x}$ assures that in the limit $\varepsilon \rightarrow 0$, the differential-algebraic equations (DAE) system that describes the slow dynamics of Eq. (\ref{eq:generalsin}) (in the next section) has a finite index and a well-defined solution, and it is satisfied in typical chemical process applications.

We obtain the set of linearly independent constraints (\ref{eq:4.40}) that must be satisfied in the slow time-scale $t$. In the fast time-scale $\tau$, the algebraic constraints Eq. (\ref{eq:4.40}) are not satisfied unless steady-state reaches \cite{baldea_dynamics_2012}.

The initial condition $x(0)=x_0$ of the system (\ref{eq:generalsin}) applies in the inner approximation i.e. the initial condition of the fast dynamics $x_f(\tau=0)=x_{f0}=x_0$. 

\subsubsection*{Decomposition of slow dynamics}

In the outer approximation, multiplying system (\ref{eq:generalsin}) by $\varepsilon$ and considering the limiting case $\varepsilon \rightarrow 0$ in the slow time-scale, we obtain the constraint $k(x_s)=0$  which includes $p_x$ linearly independent scalar equations. Note that $x_s$ denotes the vector of slow states. Note also that the constraint $k(x_s)=0$ should be satisfied by the slow dynamics for all time. This also implies that the constraint should be satisfied by the slow dynamics at time $t=0$, which may be used to determine the initial condition for the slow dynamics.

Taking the limit $\varepsilon\rightarrow 0$ and defining
\begin{equation}
    z = \lim_{ \varepsilon \rightarrow 0} \frac{k(x_s)}{\varepsilon}
    \label{eq:4.5}
\end{equation}
system (\ref{eq:generalsin}) becomes,
\begin{equation}
\begin{split}
    \frac{dx_s}{dt}&= f(x_s)+ g(x_s)u+b(x_s)z \\
      k(x_s)&=0
\end{split}
\label{eq:4.6}
\end{equation}

Note that in (\ref{eq:4.6}), $z$ is indeterminate. Once the input $u(t)$ is specified (e.g. by a control law), it is possible to differentiate the algebraic constraint in Eq. (\ref{eq:4.6}) to obtain (after differentiating a sufficient number of times depending on the index  number) a solution for the algebraic variable $z$. Without the loss of generality, it is assumed in this work that by one differentiation in time of the algebraic constraint, a solution of $z$ can be obtained. Based on this assumption, it is obtained that: 
\begin{equation}
     z=-[L_b(k(x_s))]^{-1} [L_{f}(k(x_s))+L_{g(x_s)}(k(x_s))u]\\
     \label{eq:1.7}
\end{equation}

The matrix $L_b(k(x_s))$ denotes the Lie derivative of function $k(x_s)$ along $b(x_s)$ and is nonsingular. By substituting $z$ in Eq. (\ref{eq:4.6}), we obtain an approximation of the slow dynamics of system (\ref{eq:generalsin}):
\begin{subequations}
\label{eq:4.7}
\begin{align}
  \begin{split}
      \frac{dx_s}{dt}= &f(x_s)+ g(x_s) u+b(x_s)(-[L_b(k(x_s))]^{-1} [L_{f}(k(x_s))+L_{g}(k(x_s))u]):=f_s(x_s, u)  \label{4.7ba} 
   \end{split} \\
     k(x_s)=&0 \label{4.7b}
\end{align}
  \end{subequations}
with an initial condition $x_{s0}=x_{fss}$. $f_s(x_s, u)$ is the analytic vector field of the dimension of $n_{x_s}$.

For more detailed procedures on decomposing fast and slow dynamics, the reader is referred to \cite{baldea_dynamics_2012}.

\subsubsection*{Reconstruction of the actual dynamics from fast and slow dynamics}

In the preceding discussion, system (\ref{eq:generalsin}) is brought down to a fast subsystem and a slow subsystem. The fast subsystem approximates the fast dynamics in the original system and the slow subsystem approximates the slow dynamics in the original system. We also see these explicit equations for fast and slow subsystems are each valid in their corresponding time-scale $\tau$ and $t$, respectively. An approximation of the actual dynamics of the original system can be constructed based on the fast and slow subsystems.  

We use the idea of matching to find out the overlap region of fast and slow subsystems. The overlap region is the intermediate area where both fast and slow approximations should agree for identical values. To elaborate, let us consider the fast subsystem; this approximation dominates in a certain region of its domain. Similarly, the slow subsystem dominates in a specific but distinct area of approximation. However, there is a common region where the approximations overlap. The overlap value is $x_{olp}$ which is the outer limit of the fast subsystem, or the inner limit of the slow subsystem. That is, $x_{olp}=\lim_{\tau \to \infty}x_{f}(\tau)=\lim_{t \to 0} x_{s}(t)$ \cite{verhulst_methods_2005, kevorkian_multiple_1996}.

To obtain the final matched and composite solution, valid on the whole time domain, the uniform method is one of the popular methods. It adds the inner and outer approximations and subtracts their overlap value, $x_{olp}$, which would otherwise be counted twice. Basically, the overlap value is $x_{fss}$ found from the limits mentioned above $(x_{olp}=x_{fss})$. Therefore, the final composite solution $x_{cp}$ which is applicable in the entire time $t$ domain \cite{verhulst_methods_2005, kevorkian_multiple_1996}.
\begin{equation}
\begin{split}
    x_{cp}(t)= x_{f}(\varepsilon \tau)+ x_{s}(t)- x_{fss}
\end{split}
\label{eq:4.60}
\end{equation}
When the exact solution $x$ for a singular perturbation problem Eq. (\ref{eq:generalsin}) is not available, $x_{cp}$ is an approximate solution of such system that remains uniformly valid in the independent variable $t$. 

\section*{Proposed distributed state estimation scheme}

\begin{figure}[t]
	\centering
	\includegraphics[scale=0.7]{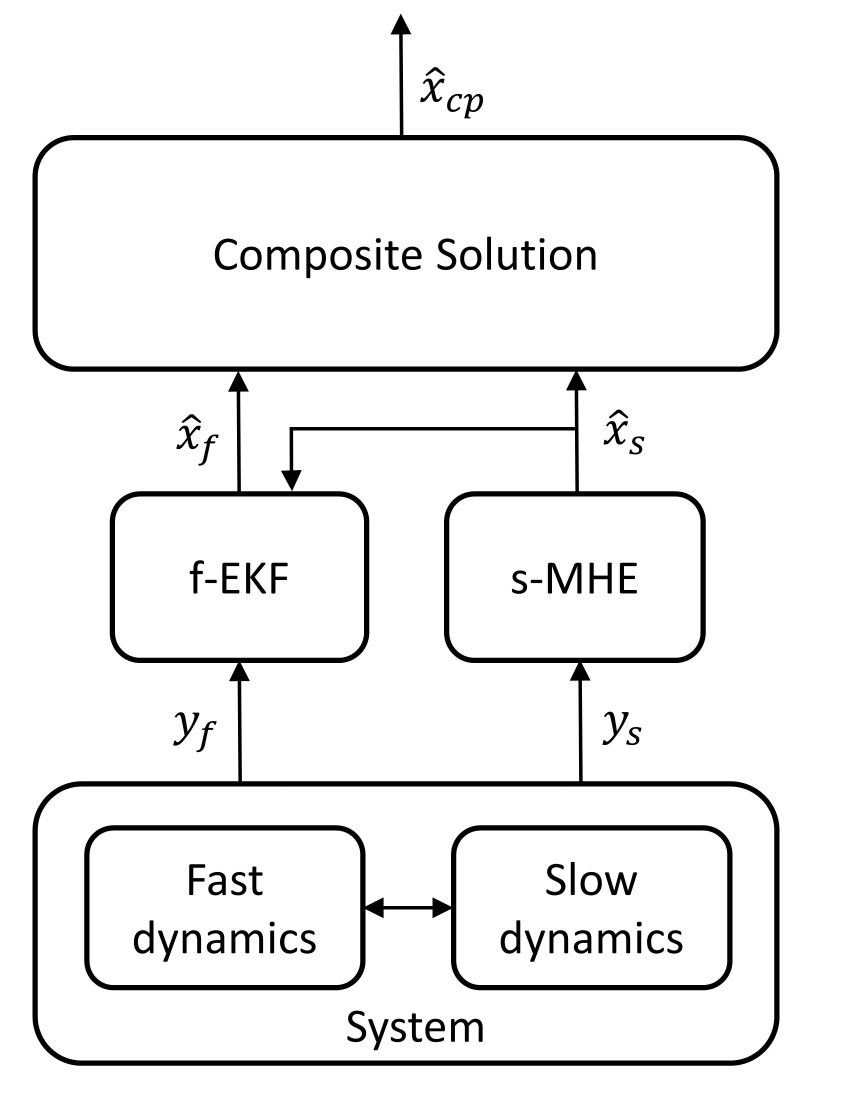}
	\caption{A schematic of the proposed distributed state estimation scheme}
	\label{fig:scheme}
\end{figure}

In this section, we propose a distributed state estimation scheme to estimate the state of the two-time-scale system (\ref{eq:generalsin}) based on
fast and slow dynamics decomposition. A schematic of the proposed distributed state estimation scheme is presented in Fig. \ref{fig:scheme}. A local estimator is designed for each fast subsystem and slow subsystem. Two different estimators are used: extended Kalman filter (EKF) is designed for the fast subsystem, and moving horizon estimation (MHE) is associated with the slow subsystem. The reason to use two different estimators is mainly the existing different time-scales. EKF takes less time to evaluate, which is apt for the fast subsystem, but it cannot take nonlinearity or constraints into account in a systematical way and may give poor performance. The use of EKF for the fast dynamics is a tradeoff between computing speed and performance. On the other hand, MHE is more suitable for complex nonlinear and constrained dynamic systems. But, it requires online solutions of dynamic optimization problems, which results in increased computational cost. The sluggish nature of the slow subsystem and the necessary high accuracy in exchange for an increase in computational cost are reasonable enough for the consideration. The EKF and MHE are designed based on the reduced fast subsystem and the reduced slow subsystem derived in the previous section. Note that f-EKF denotes EKF for the fast subsystem, and s-MHE indicates MHE for the slow subsystem. 

There is no information exchange between the f-EKF and s-MHE. Since each subsystem evolves at different time-scales, it is desirable to use different sampling periods in the local estimator designs for the fast and slow subsystems. Therefore, the sampling period for f-EKF and s-MHE are defined as $\Delta_f$ and $\Delta_s$ respectively. Without loss of generality, we assume that $\Delta_s$ is integer multiple of $\Delta_f$, i.e., $\Delta_s = n\Delta_f$ where $n$ is a positive integer. In the proposed design, we use $\tau_q := \tau_0 + q \Delta_f$ with $q \geq 0$ and $t_k :=t_0 + k \Delta_s$ with $k \geq 0$ to denote the sampling instants of f-EKF and s-MHE, respectively. While $\hat{x}_f(\tau_q)$ denotes the state estimates of f-EKF at $\tau_q$, $\hat{x}_s(t_k)$ is the state estimates of s-MHE at $t_k$. We denote $y_f(\tau_q)$ and $y_s(t_k)$ as the measurements for f-EKF and s-MHE sampled at $\tau_q$ and $t_k$, respectively. In the end, we find the estimation of the actual state based on composite solution Eq. (\ref{eq:4.60}). In the following discussion, we illustrate the proposed estimator design procedure that accounts rationally for the nonlinear two-time-scale dynamics. 

\subsection*{Proposed implementation algorithms}

We decompose two-time-scale systems described in Eq. (\ref{eq:generalsin}) into two separate reduced subsystems evolving in a fast and a slow time-scales as illustrated in the previous section. The fast subsystem is described by Eq. (\ref{eq:1.4}) and the slow subsystem is described by Eq. (\ref{eq:4.7}).

It is important to note that the measurements used in the f-EKF and s-MHE are directly obtained from the actual system measurement $y$ of Eq. (\ref{eq:1b}) but sampled every $\Delta_f$ and $\Delta_s$ respectively. 

In the proposed scheme, the f-EKF and s-MHE are designed independently based on the above subsystems. The implementation details of the distributed state estimation are specified in the following algorithm:
\begin{algorithm}
\caption{Proposed estimation algorithm}
\begin{algorithmic}[1]
\Initialization Initialize the f-EKF and s-MHE with their initial guesses. Find steady state of the reduced fast subsystem model
\FOR{$q = 0,1,2,3 \dots $}
\State At $\tau_q$, receive measurement $y_f(q)$
\State Evaluate the f-EKF to obtain $\hat x_{f}(\tau_q)$ 
\IF{$\frac{q}{n} \text{is an integer}$}
\STATE Evaluate the s-MHE to obtain $\hat x_{s}(t_k)$ and send $\hat x_s(t_k)$ to the f-EKF
\ELSE
\STATE Obtain open loop prediction $\hat x_s$ from the reduced slow subsystem model and send $\hat x_s(\tau_q)$ to the f-EKF
\ENDIF
\STATE Compute $\hat{x}_{cp}(t)$ at time instant $\tau_q$
\ENDFOR
\end{algorithmic}
\end{algorithm}

\subsection*{Design of f-EKF}

In this section, we design a EKF estimator based on the reduced fast subsystem model to estimate the state of the fast subsystem. Specifically, in the design of the EKF, we consider the fast subsystem with additive process noise and the system output represented in terms of the fast and slow states based on (\ref{eq:4.60}) as follows:
\begin{subequations}
\begin{align}
     \frac{dx_f}{d\tau}&=f_f(x_f)+w_f \label{eq:1.20a}\\
  y(\tau)&=h(x_f+x_s-x_{fss})+v \label{eq:1.20b}
\end{align}
\label{eq:1.20}
\end{subequations}
\vspace{-3mm}

Note that in (\ref{eq:1.20a}), $w_f$ reflects the modeling error of the subsystem model. The modeling error may come from the system disturbance of the original system ($w$ in (\ref{eq:1a})) and the assumption of $\varepsilon=0$ in deriving the fast subsystem. (\ref{eq:1.20b}) implies that in the design of the f-EKF,
information of the slow subsystem state $x_s$ and the steady-state information of the fast subsystem $x_{fss}$ is needed. 

EKF is a common method used for state estimation of nonlinear systems based on successively linearizing the nonlinear system. It can be divided into two steps, which are prediction and update steps \cite{yin_state_2018}.

\paragraph{Prediction step.} At a sampling time $\tau_{q-1}$, $q=1,2,\ldots$, in an open-loop manner based on the fast subsystem model and the estimate of the fast subsystem state at $\tau_{q-1}$, the f-EKF first predict the state at the next sampling time.
\begin{align}
{\hat{x}_f(\tau|\tau_{q-1})}={\hat{x}_f(\tau_{q-1}|\tau_{q-1})}+ \int_{\tau_{q-1}}^{\tau_{q}} f_f(\hat{x}_f\left( \tau|\tau_{q-1}) \right) d\tau
\end{align}
where ${\hat{x}_f(\tau|\tau_{q-1})}$ represents the prediction of
the state at time instant $\tau \in (\tau, \tau_{q-1}]$. The propagation of the process disturbance is as follows: 
\begin{align}
\dot{P}_f(\tau|\tau_{q-1})= F_f(\tau,\tau_{q-1})P_f(\tau_{q-1}|\tau_{q-1})F_f^T(\tau,\tau_{q-1})+\int_{\tau_{q-1}}^{\tau} F_f(\tau,t) Q_f F_f^T(\tau,t) dt
\end{align}
where $P_f$ and $Q_f$ are the error covariance matrix and the state covariance matrix, respectively, $P(\tau_q|\tau_{q-1})$ is a square matrix containing the a \textit{priori} estimation error covariance information, and $F_f(\tau,\tau_{q-1})$ denotes the state transition matrix of the time-varying linearized system matrix, $A_f(\tau|\tau_{q-1}) :=\frac{\partial{f_f}}{\partial{x_f}} |_{(\hat{x}_f(\tau|\tau_{q-1}))}$ and can be calculated as follows:
\begin{align*}
   \frac{\partial F_f(\tau,\tau_{q-1})}{\partial \tau}&=A_f(\tau|\tau_{q-1}) F_f(\tau,\tau_{q-1})\\
   \text{s.t.} \hspace{2mm} F_f(\tau,\tau)&= I 
\end{align*}
for $\tau \in [\tau_{q-1}, \tau_q]$ with $I$ being the identity matrix.

\paragraph{Update step.} At each sampling instant $\tau_q$, a state estimate of the actual
dynamics of the fast subsystem (denoted as $\hat x_f(\tau_q|\tau_q)$) is obtained by performing the measurement-update step. $K(\tau_q)$ is the correction gain updated at $\tau_q$
which is used to minimize a \textit{posteriori} error covariance based
on the measurement innovation (i.e. $y_f(\tau_q)-Ch(\hat{x}_f(\tau_q|\tau_{q-1})+ \hat{x}_s-x_{fss})$).
\begin{align}
    K(\tau_q)=P_f({\tau_q|\tau_{q-1}})H^T(\tau_q)(H(\tau_q)P_{q|q-1}H^T(\tau_q)+R_{f})^{-1}
\end{align}
where $H(\tau_q) = \frac{\partial{h}}{\partial{x_f}} |_{(\hat{x}_{f}(\tau_q|\tau_{q-1}))}$ is the observation matrix, $R_f$ is the covariance matrix of the measurement noise $v_f$. The updated state estimate is as follows:
\begin{align}
    \hat{x}_f(\tau_q|\tau_q)=\hat{x}_f(\tau_{q-1}|\tau_{q-1})+K(\tau_q) \left ( y_f(\tau_q)-h(\hat{x}_f(\tau_q|\tau_{q-1})+ \hat{x}_s-x_{fss}) \right)
    \label{eq:14}
\end{align}
where $\hat{x}_f(\tau_q|\tau_q)$ represents the estimate of $x_f$ at time $\tau_q$ given observations up to time $\tau_q$. The updated state covariance is as follows:
\begin{align}
    P_{f}(\tau_q|\tau_q)= (I-K(\tau_q)H(\tau_q))P_{f}(\tau_q|\tau_{q-1})
\end{align}
where $P(\tau_q|\tau_q)$ is the a \textit{posteriori} error covariance matrix with respect to the estimation error at $\tau_q$, and $I$ is the identity matrix with dimension $n_{x_f}$. Note that in the above f-EKF design, $R_f$, $Q_f$, $P_f$ are three tuning parameters. 

\subsection*{Design of s-MHE}

In this section, we design the s-MHE based on the reduced slow subsystem model to estimate the slow states. Similarly, we consider a stochastic version of the reduced slow subsystem model in Eq. (\ref{eq:4.7}) described as in the following form:
\begin{subequations}
\begin{align}
\begin{split}
     \frac{dx_s}{dt}=&f(x_s)+ g(x_s) u+b(x_s)(-[L_b(k(x_s))]^{-1} [L_{f}(k(x_s))+L_{g}(k(x_s))u])+w_s \label{eq:1.10a}
\end{split}\\
      \begin{split}
          k(x_s)=&0 \label{eq:1.10b}
      \end{split}\\
    \begin{split}
        y(t)=&h(x_s)+v \label{eq:1.10c}
    \end{split}
\end{align}
\label{eq:1.10}
\end{subequations}\vspace{-9mm}

In (\ref{eq:1.10a}), $w_s$ accounts the modeling error of this subsystem model which may originate either from the actual system ($w$ in (\ref{eq:1a})) or the assumption for decomposition of slow subsystem. As states earlier, when deriving the slow subsystem, it is assumed that the fast dynamics have converged to the corresponding steady state values. Based on the assumption and the expression shown in (\ref{eq:4.60}), the output equation (\ref{eq:1.10}) can be obtained for the slow subsystem.

MHE is an online optimization based estimation method \cite{rao_constrained_2003}. The proposed s-MHE optimization problem at time $t_k$ is formulated as follows:
\begin{subequations}
\begin{align}
    \min\limits_{x_s(k-N), \hat{w}_s(\cdot)} &  
||\hat x_s(k-N) - \tilde{x}_s(k-N)||^2_{P_s^{-1}} +
\sum\limits_{j=k-N}^{k-1}||\hat{w}_s(j)||^2_{Q_s^{-1}} + \sum\limits_{j=k-N}^{k}||\hat{v}(j)||^2_{R_s^{-1}}
\vspace{2mm}
\label{17a}\\
{\rm s.t.~} &   \hat x_s(j+1) = f_s(\hat x_s(j),u(j))+\hat w_s(j),~~ j \in [k-N, k-1] \vspace{2mm} 
\label{17b}\\
& \hat{v}(j) = y(j) - h(\hat x_s(j)), ~~ j \in [k-N, k]\vspace{2mm} 
\label{17c}\\
& \tilde{x}_s(k-N) =\hat x_s(k-N|k-N) \vspace{2mm}
\label{17d}\\
& x_s \in \m X_s , \hat{w}_s\in \m W, \hat{v} \in \m V 
\label{17e}
\end{align}
\end{subequations}
where $\hat x_s$ denotes the estimated value of the slow subsystem state $x_s$, $\hat w_s$ denotes the
estimated system disturbance, $\hat v$ denotes the estimated
measurement noise, and $\m X_s$, $\m W$, and $\m V$ denote the known constraints on the augmented state, the system disturbance, and the measurement noise, respectively.
Equation (\ref{17a}) is the cost function the MHE tries to minimize. The objective of the s-MHE is to find the best estimates of the system states such that the model disturbance and measurement noise are minimized. $P_s^{-1}$, $Q_s^{-1}$ and $R_s^{-1}$ are positive definite weighting matrices which are tuning parameters. The arrival cost, $||\hat x_s- \tilde{x}_s||^2_{P^{-1}_s}$ summarizes the information from the initial state of the model up to the beginning of the estimation window of the MHE. $N$ denotes the length of the estimation window. Equations (\ref{17b}) and (\ref{17c}) are the slow subsystem model with system disturbance and measurement noise considered. In Eq. (\ref{17d}), $\hat x_s(k-N|k-N)$ represents
the estimated state $\hat x_s$ at time instant $k - N$, which is estimated at time instant $k - N$. Equation (\ref{17e}) is the known constraints or compact sets that bound the subsystem state, system disturbance, and measurement noise. 

\section*{Application to a CSTR}

\subsection*{Process description}

\begin{figure}
\begin{center}
{
  \includegraphics[width=0.4\textwidth]{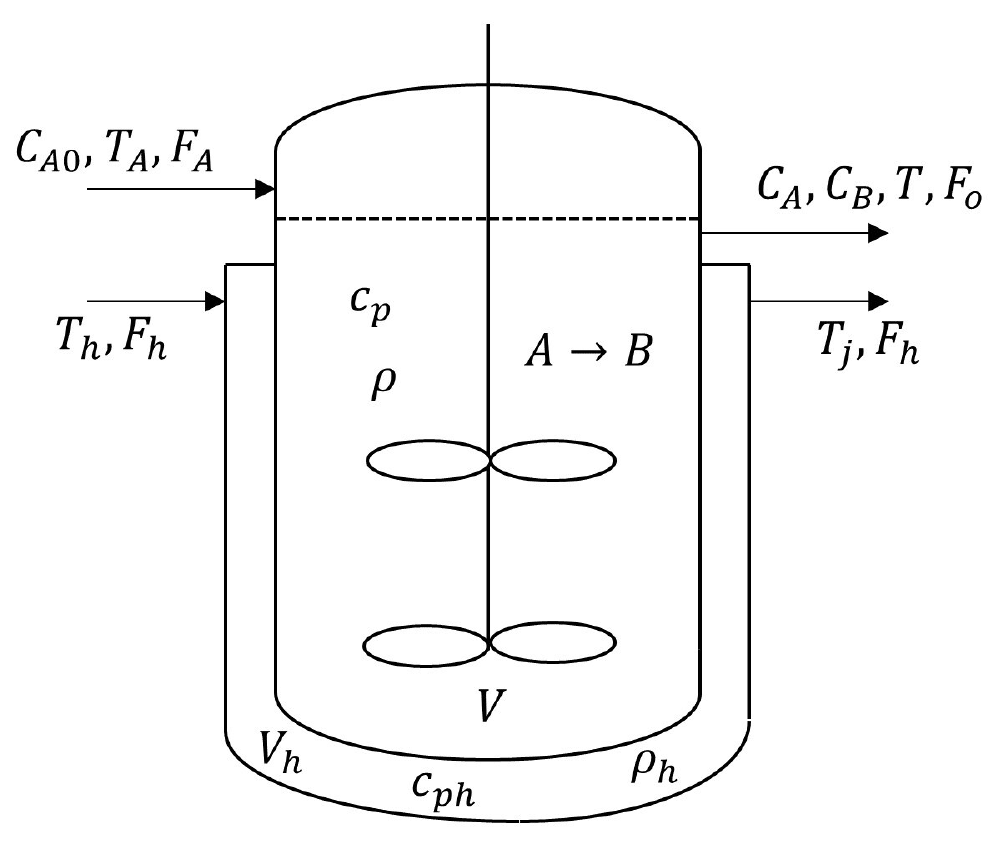}
 } 
 	\caption{A continuous-stirred tank reactor with heating
jacket}
	\label{fig:1}
\end{center}
\end{figure}

Consider a continuous-stirred tank reactor (CSTR) with heating jacket in Figure \ref{fig:1}. Reactant A is fed to the reactor at a flowrate $F_A$, initial molar concentration $C_{A0}$ and temperature $T_A$. The reactant A is converted into the product B through the irreversible endothermic reaction $A\xrightarrow{}B$, and the product stream is withdrawn at a flowrate $F_0=F_A$. This implies that the reactor holdup volume $V$ is constant. The reaction rate $r_A$, is given by the following Arrhenius expression:
\begin{align}
    r_A=k_0 \exp{ \left( \frac{-E}{RT} \right) } C_A V
\end{align}
 where $k_0$ and $E$ are the reaction rate coefficient and activation energy, respectively, $T$ is the reactor temperature, and $C_A$ is the molar concentration of A in the reactor. Heat is provided to the reactor from the jacket, where a heating fluid is fed at a flowrate $F_h$ and a temperature $T_j$. The modeling equations for the process include the mole balances for the two components in the reactor and the energy balances in the reactor and the jacket. The resulting dynamic model is as follows:\\
\begin{equation}
\begin{split}
    &\dot{C_A} = \frac{F_A}{V}(C_{A0}-C_A)-r_A\\
  & \dot{C_B} = -\frac{F_A}{V}C_B+r_A\\
  & \dot{T}   =\frac{F_A}{V}(T_A-T)-r_A \frac{\Delta H_r}{\rho c_p}+\frac{UA}{\rho c_p} \left (\frac{T_j-T}{V} \right)\\
  & \dot{T_j} = \frac{F_h}{V_h}(T_h-T)-\frac{UA}{\rho_h c_{ph}}\left(\frac{T_j-T}{V_h}\right)
     \end{split}
     \label{cstr1}
\end{equation}
where, $c_p$ and $c_{ph}$ are the specific heat capacities of the reaction mixture in the reactor and heating liquid in the jacket, respectively. Similarly, the density of the liquids of the reactor and jacket are $\rho$ and $\rho_h$, respectively. $U$ is the overall conductance or heat transfer coefficient, $A$ is the heat transfer area of the contact surface between ractor and jacket, and $\Delta H_r$ is the heat of reaction which is the enthalpy of the reaction.

It is assumed that the density and specific heat capacities
of the two liquids are the same, i.e. $c_p=c_{ph}$ and $\rho=\rho_h$ and the liquid holdup in the jacket at a temperature $T_j$ has a constant volume $V_h$. The heat transfer rate by convection thought the contact surface can be expressed as:
\begin{align}
    \dot{Q}=UA(T_A-T)
\end{align}
Furthermore, we assumed that the heat transfer between the heating jacket and the reactor is fast compared to the reaction occurring in the reactor. The large difference in heat transfer and reaction in the model induces multiple-time-scale behavior in this dynamical systems. Therefore the ratio between the heat transfer to the reaction is defined as:
\begin{align}
    \frac{1}{\varepsilon}=\frac{UA}{\rho c_p}
\end{align}
where, $\varepsilon$ is a small parameter or singular perturbation parameter which ensures the presence of fast and slow transients in time response of the system. Based on the definition $\varepsilon$, the CSTR model (\ref{cstr1}) can be rewritten as follows:\\
\begin{align}
\begin{split}
        \dot{C_A}&= \frac{F_A}{V}(C_{A0}-C_A)-r_A\\
    \dot{C_B}&= -\frac{F_A}{V}C_B+r_A\\
    \dot{T}&=\frac{F_A}{V}(T_A-T)-r_A \frac{\Delta H_r}{\rho c_p}+\frac{1}{\varepsilon}\left(\frac{T_j-T}{V}\right)\\
    \dot{T_j}&= \frac{F_h}{V_h}(T_h-T)-\frac{1}{\varepsilon} \left( \frac{T_j-T}{V_h}\right)
\end{split}
\label{eq:19}
\end{align}

For this process, it is considered that the state vector is $x=[x_1,\hspace{1mm}x_2, \hspace{1mm}x_3,\hspace{1mm} x_4]^T=[C_A, \hspace{1mm} C_B,\hspace{1mm} T,\hspace{1mm} T_j]^T$, the manipulated input vector is $u=[u_1,\hspace{1mm} u_2]^T=[F_A,\hspace{1mm} F_h]^T$, and the controlled output vector is $[y_1, \hspace{1mm} y_2]^T= [x_2, \hspace{1mm} x_4]^T$. The model in (\ref{eq:19}) takes the form of Eq. (\ref{eq:1a}) with the system functions defined as\\
\\
$f(x)=\begin{bmatrix} 
k_0 \exp{\left (\displaystyle \frac{-E}{Rx_3} \right)} x_1 \\
\\
-k_0 \exp{\left (\displaystyle\frac{-E}{Rx_3}\right)} x_1 \\
\\
-k_0 \exp{\left (\displaystyle\frac{-E}{Rx_3}\right)} x_1 \displaystyle \frac{\Delta H_r}{\rho c_p}\\
\\
0\\
\end{bmatrix}$ \hspace{2mm}
$g(x)=\begin{bmatrix}
\displaystyle \frac{C_{A0}- x_1}{V} & 0\\
\\
\displaystyle \frac{-x_2}{V} & 0 \\
\\
\displaystyle \frac{T_A- x_3}{V} & 0\\
\\
0 & \displaystyle \frac{T_h- x_4}{V_h}\\
\end{bmatrix}$\hspace{2mm}
$b(x)=\begin{bmatrix}
0\\
\\
0 \\
\\
\displaystyle \frac{1}{V}\\
\\
-\displaystyle \frac{1}{V_h}\\
\end{bmatrix}$\hspace{2mm}
\\
\\
$k(x)=\begin{bmatrix}
 x_4-x_3
\end{bmatrix}$

\subsection*{Subsystem decomposition }

Following the method described in section 2.3.1, the fast dynamics of the process can be obtained and the fast subsystem is shown below:
\begin{align}
\begin{split}
    \frac{dx_{3f}}{d\tau} &=\frac{x_{4f}-x_{3f}}{V}\\
    \frac{dx_{4f}}{d\tau} &=-\frac{x_{4f}-x_{3f}}{V_h}
    \end{split}
    \label{fasteq}
\end{align}
According to the developments in Section 2.3.2, we multiply Eq. (\ref{eq:19}) by $\varepsilon$ and consider the limit of an infinitely high heat transfer rate compared to reaction $(\varepsilon \xrightarrow{} 0)$ in the original time-scale $t$. In this limiting case, the heat transfer resistance becomes negligible,
and the reactor and jacket approach thermal equilibrium. The heat transfer rate $\dot{Q}$ is driven by the thermal equilibrium condition $ x_{4s} \rightarrow{} x_{3s}$ instead of the explicit heat transfer correlation. We obtain the constraints $k(x_s)$ which is the linearly independent constraints,
\begin{align}
    x_{4s}-x_{3s}=0
    \label{eq:24}
\end{align}
which must be satisfied in the slow time-scale. Also in the limit $(\varepsilon \xrightarrow{} 0)$, the term $\frac{( x_{4s}-x_{3s})}{\varepsilon}$ which corresponds to the differences of large heat transfer present in the energy balance equations become indeterminate. Therefore, $z= \lim_{ \varepsilon \rightarrow 0} \frac{( x_{4s}-x_{3s})}{\varepsilon}$ is defined as the algebraic variable, the vector of the finite, but unknown term. The slow dynamics becomes,
\begin{align}
\begin{split}
    \dot{x}_{1s} &= \frac{F_A}{V}(C_{A0}-x_{1s})-k_0 \exp\left ({\frac{-E}{R x_{3s}}}\right) x_{1s} V   \\
    \dot x_{2s} &= -\frac{F_A}{V}x_{2s}+ k_0 \exp\left ({\frac{-E}{R x_{3s}}}\right) x_{1s} V\\
    \dot{x_{3s}}   &=\frac{F_A}{V}(T_A-x_{3s})-k_0 \exp\left ({\frac{-E}{Rx_{3s}}}\right) x_{1s} V \frac{\Delta H_r}{\rho c_p}+\frac{z}{V}\\
    \dot{x_{4s}} &= \frac{F_h}{V_h}(T_h-x_{4s})-\frac{z}{V_h} 
\end{split}
 \end{align}
which represents the model of the slow dynamics of the process. Thus, the $z$ variable can be obtained after just one differentiation of the algebraic constraints from Eq. (\ref{eq:24}). The values of the process parameters and variables at the nominal steady state are given in Table \ref{t1}.
 
Corresponding to the parameter values shown in Table \ref{t1}, the process has a
steady-state \\$[C_A,\hspace{1mm} C_B, \hspace{1mm}T, \hspace{1mm}T_j]^T=[1.205 \hspace{2mm} \text{mol/l},\hspace{1mm} 1.295 \hspace{1mm} \text{mol/l}, \hspace{1mm} 302.3\hspace{1mm} \text{K}, \hspace{1mm} 302.6\hspace{1mm} \text{K}]^T$ when input of the system is $[F_A, \hspace{1mm}F_h]^T=[2.0 \hspace{1mm} \text{l/s} , \hspace{1mm} 0.1 \hspace{1mm} \text{l/s}]^T$. It was verified that these values correspond to a stable steady state of system (\ref{eq:19}).

\subsection*{f-EKF and s-MHE designs}

In this section, we take advantage of the configured subsystems equations and implement a distributed state estimation scheme for the CSTR. Within the proposed distributed framework, two local estimators are designed for the two subsystems, an EKF is designed for the fast subsystem, while an estimator is developed based on nonlinear MHE for the slow subsystem. At every sampling instant, each local estimator is evaluated to provide subsystem state estimates and those are used to reconstruct the actual state estimate.

In the design of the state estimators, it is assumed that $C_B$ and $T_j$ are the measurements of the system and the objective is to estimate the entire state vector of the system based on the two outputs using the proposed distributed state estimation scheme. In the simulations, the random process disturbance is generated following a normal distribution with zero mean and standard deviation 0.1. Similarly, the random measurement noise is considered to be Gaussian white noise with mean zero and standard deviation 0.001. The weighting matrices are diagonal matrices such as $Q_f=10^{-2}\text{diag}([1, \hspace{2mm} 1])$, $R_f=10^{-6}\text{diag}([1, \hspace{2mm} 1])$ and $P_f=10^{-8}\text{diag}([1, \hspace{2mm} 1])$ for the f-EKF. For the design of s-MHE, the estimation window size is selected to be $3$. The weighting matrices for s-MHE are $Q_s=10^{-2}\text{diag}([1, \hspace{2mm} 1, \hspace{2mm} 1, \hspace{2mm} 1])$, $R_s=10^{-6}\text{diag}([1, \hspace{2mm} 1])$ and $P_s=10^{-8}\text{diag}([1, \hspace{2mm} 1,  \hspace{2mm} 1,  \hspace{2mm} 1])$. The f-EKF is evaluated at a fast sampling time $\Delta_f = 0.01 \hspace{2mm}\text{s}$  while the s-MHE is evaluated at a slow sampling time $\Delta_s = 0.1 \hspace{2mm}\text{s}$. 

Next, we implement the proposed distributed scheme to the CSTR process to investigate the performance of f-EKF and s-MHE designs. Also, to compare its performance, we introduce two different state estimation schemes. First, considering the decomposition, we design a decentralised estimator scheme based on each subsystems. The s-MHE does not send any information to the f-EKF. This design follows the same sets of equations derived for f-EKF except Eq. (\ref{eq:14}). It takes only current predicted fast state estimates $\hat{x}_f(\tau_q|\tau_{q-1})$ to update $\hat{x}_f(\tau_q|\tau_{q})$ so there is no information exchange between the estimators. The covariance matrices for f-EKF and s-MHE are kept the same as the distributed scheme. Furthermore, we design a centralized MHE estimator for the actual system. The centralized scheme is established based on Eq. (\ref{cstr1}) before decomposition of the system. The weighting matrices for the centralized MHE are $Q=10^{-2}\text{diag}([1, \hspace{2mm} 1, \hspace{2mm} 1, \hspace{2mm} 1])$ and $R=10^{-6}\text{diag}([1, \hspace{2mm} 1])$ and $P=10^{-8}\text{diag}([1, \hspace{2mm} 1,  \hspace{2mm} 1,  \hspace{2mm} 1])$. And, the sampling time and estimation window length are considered to be 0.01 s and 3, respectively. For all these schemes, the system disturbance and measurement noise are the same for proper comparison.
 
 In the following simulations, a couple of indexes are used to evaluate the performance of the estimators. The average relative standard deviation $\sigma_{x_i}$ is defined as
\begin{align}
    \sigma_{x_i}= \sqrt{\dfrac{1}{N_{sim}}\sum\limits_{j=0}^{N_{sim}-1}\left (\frac{\hat x_i(t_j)-x_i(t_j)}{x_i(t_j)}\right)^2}
\end{align}
where $N_{sim}$ indicates the total simulation steps, $\hat x_i$ denotes the estimated value, and $x_i$ denotes the actual value of the $i^{th}$ state for $i=1, 2,...4$. Another performance indexes is the average root-mean-square error (RMSE) over the time period
\begin{align}
    \text{RMSE}=\dfrac{1}{N_{sim}} \sum\limits_{j=0}^{N_{sim}} \sqrt{\dfrac{1}{4}\sum\limits_{i=1}^{4}\left( \frac{\hat x_i(t_j)-x_i(t_j)}{x_i(t_j)}\right)^2}
\end{align}
These performance indexes are expressed in percentage to evaluate the performance of each scheme.

\subsection*{Simulation results}

In this section, the effectiveness of the proposed two time-scale decomposition is investigated and the proposed distributed estimation scheme is applied to the CSTR and is compared with the mentioned decentralized and centralized estimation schemes.

\begin{figure}
\begin{center}
{
  \includegraphics[width=0.65\textwidth]{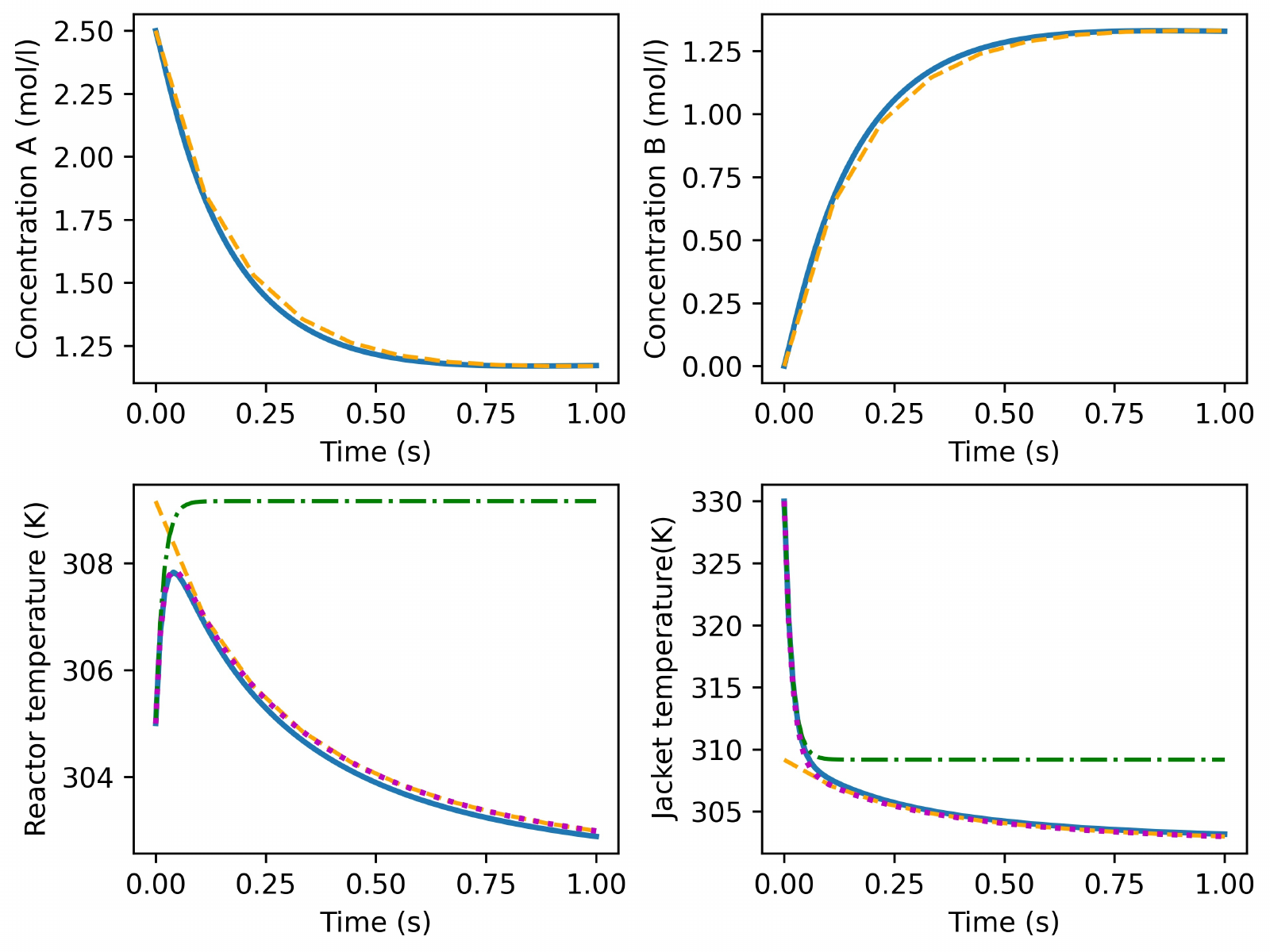}
 } 
 	\caption{Trajectories of the actual process states (blue solid lines), fast subsystem (green dot-dashed lines) and slow subsystem (orange dashed lines) and composite solution (red dotted lines)(For interpretation of the references to color in this figure legend, the reader may refer to the web version of this article.)}
	\label{fig:result1}
	\end{center}
\end{figure}

First, the effectiveness of the decomposition is investigated by comparing the composite state trajectories with the actual system state trajectories. Figure \ref{fig:result1} shows one set of the trajectories. 
The  actual system trajectory is solved using an initial condition $[2.5, \hspace{2mm} 0.0,\hspace{2mm} 305,\hspace{2mm} 330]$ for the states $C_A$, $C_B$, $T$, and $T_j$ respectively. For the composite solution, the fast subsystem is integrated with an initial condition $[305,\hspace{2mm} 330]$ for fast states $T$, and $T_j$ respectively and, the corresponding steady-state solution is found to be $[309.167, \hspace{2mm} 309.167]$. Then, the slow subsystem is solved where the initial condition is $[2.5, \hspace{2mm} 0.0,\hspace{2mm} 309.167, \hspace{2mm} 309.167]$ for $C_A$, $C_B$, $T$, and $T_j$ respectively. Due to using different sampling sizes for the fast and the slow subsystems, the slow state variables are interpolated using cubic interpolation to make compatible with the solutions of the fast subsystem. Then, according to Eq.(\ref{eq:4.60}), the composite solution is evaluated. We see that the proposed decomposition method is able to track the actual state trajectories of the chemical process. The state approximations using the composite solution are very close to the true value obtained from actual model equations. We clearly observe that the concentration trajectories are smooth and have a flat slope but the temperature trends have a very steep slope initially, and then flatten gradually. Clearly, the temperatures exhibit two time-scale behaviours existing in different time-scales. The results also show that there is no explicit separation of fast and slow variables in the system. 
It is observed that the actual system is not able to converge when the sampling time is greater than $0.04$ s. The possible reason is that the actual system is a stiff system for existing both fast and slow dynamics so it lacks numerical stability beyond step size 0.04 s. It is also seen that the composite solution starts deviating from the actual trajectories on the increase of sampling time for fast or slow subsystems. The $\sigma_{x_i}$ percentage of error for each state $C_A$, $C_B$, $T$, and $T_j$ are $0.44$, $4.022$, $1.8\times 10^{-2}$ , and $0.15$ respectively. The aim of using the subsystem decomposition method is to approximate the actual system based on the assumption $\varepsilon \rightarrow 0$ i.e., $\varepsilon$ should be sufficiently small. But in the CSTR $\varepsilon$ is the ratio of two internal properties and is 0.1 in the simulation. The use of such approximation to present the actual system leads to model mismatch. The model mismatch in terms of RMSE is 0.035\% which shows the decomposed subsystems provide very accurate approximations of the actual system.

\begin{figure}
\begin{center}
{
  \includegraphics[width=0.65\textwidth]{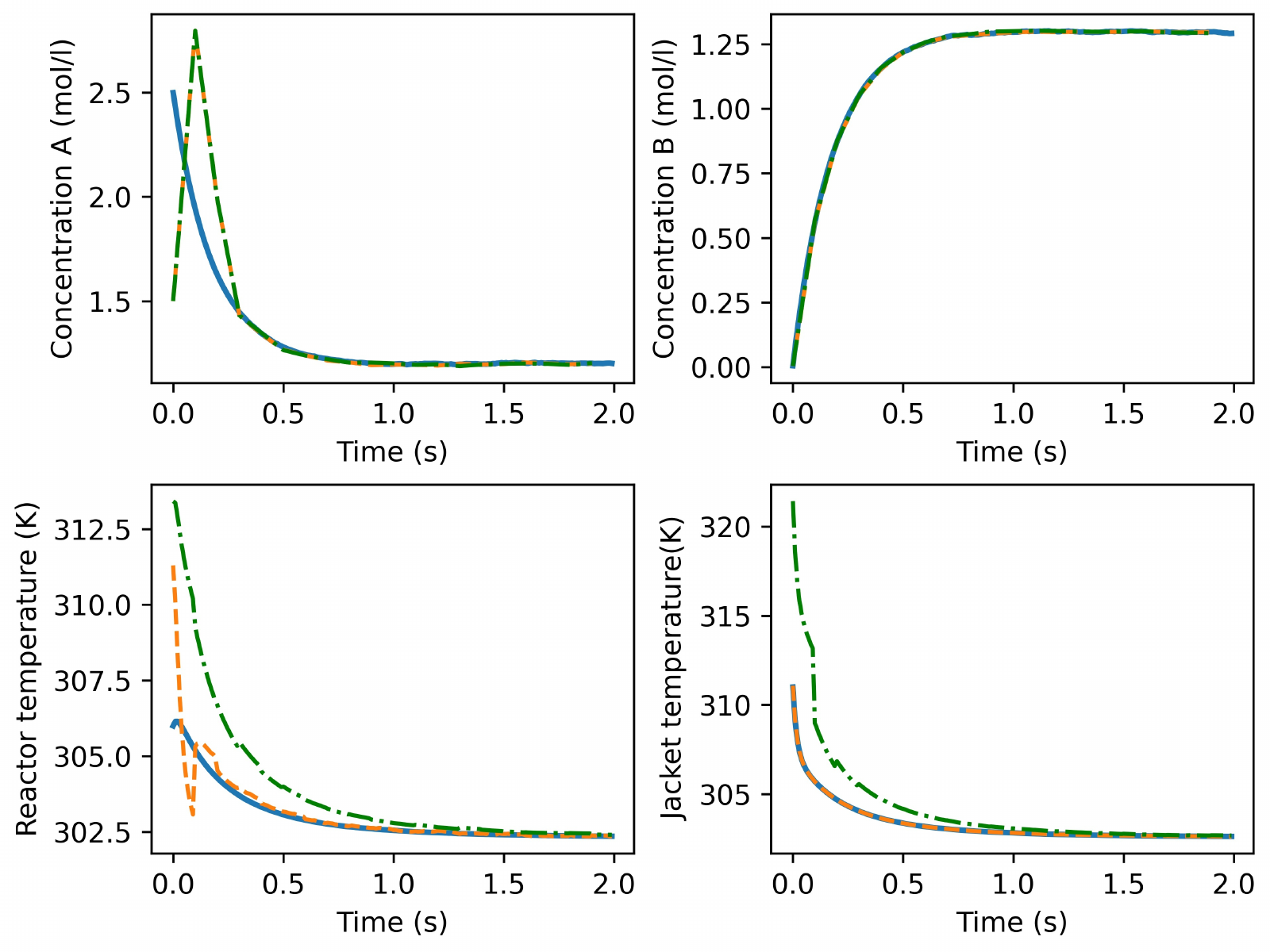}
 } 
 	\caption{Trajectories of the actual states (blue solid line), state estimates based on decomposition under proposed distributed configuration (orange dashed lines), state estimates based on decomposition under decentralized configuration (green dash-dotted lines) (For interpretation of the references to color in this figure legend, the reader may refer to the web version of this article.)}
	\label{fig:result2}
	\end{center}
\end{figure}

Next, the performance of the proposed distributed state estimation scheme is compared with other state estimation methods. Specifically, we consider three different schemes: (I) the proposed distributed state estimation; (II) f-EKF and s-MHE in a decentralized configuration; (III) a centralized MHE scheme.
We compare state estimation scheme I and scheme II presented in Figure \ref{fig:result2}. We observe that the states are able to converge the actual state trajectories in both the scheme. In Table \ref{t3}, the results show that the estimation accuracy given by the distributed MHE scheme I is better than the decentralized scheme II. Also, the maximum error of scheme II is much higher that of scheme I. It is observed that the f-EKF has improved innovation in updating the measurements because of info exchange in scheme I. Although, the simulation times are comparable, in scheme I, the states have faster convergence than scheme II. Therefore, it can be more favorable to take scheme I for state estimation for the decomposition considered. 
 
\begin{figure}
\begin{center}
{
  \includegraphics[width=0.65\textwidth]{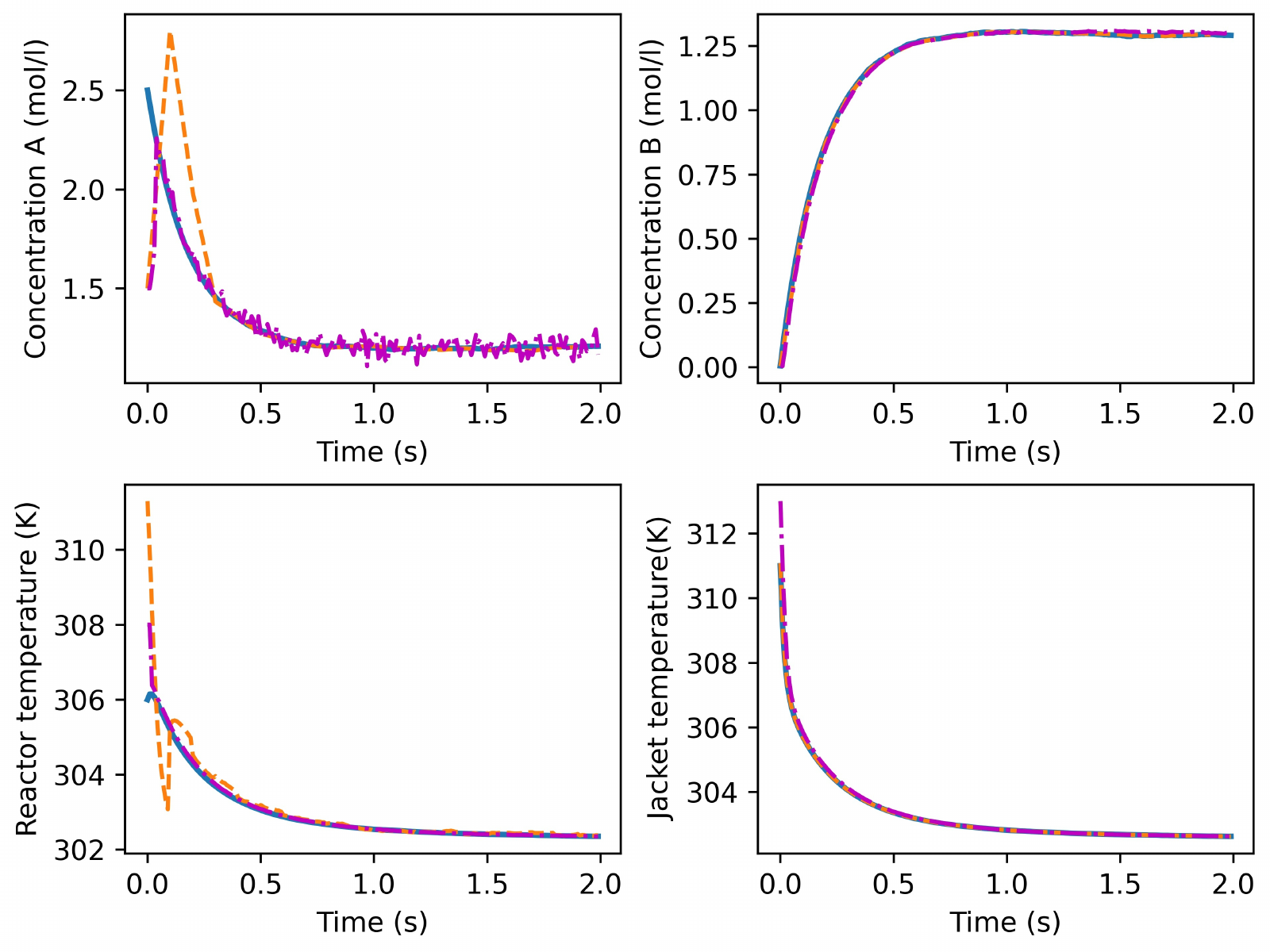}
 } 
 	\caption{Trajectories of the actual states (blue solid line), state estimates based on decomposition under distributed configuration (orange dashed lines), state estimates of actual system without decomposition (maroon dash-dotted line) (For interpretation of the references to color in this figure legend, the reader may refer to the web version of this article.)}
	\label{fig:result3}
	\end{center}
\end{figure}

Then, we consider state estimation scheme I and scheme III presented in Figure \ref{fig:result3}. The state estimators are able to track the actual state trajectories in these two schemes. But, there are potential advantages and disadvantages in applying each of the methods. It is seen that the actual system based MHE scheme III gives improved estimation performance than the decomposition based distributed scheme II. Although, Table \ref{t3} shows low $\sigma_{x_i}$ for scheme III, it takes considerably high simulation time compared to other schemes. Therefore, the performance improvement is at the cost of much higher communications and computational burdens. Another downside of the method is the sampling time for estimator should be less than 0.04 s. As the actual system can't handle the sampling time greater than 0.04 s the optimiser of MHE encounters serious numerical instability in estimating the states of the system. 

Furthermore, we conduct simulations to demonstrate the less dependence of the scheme I on the size of the estimation horizon compared to the counterpart based on regular MHE (Scheme III). We notice that the ability to use different sampling periods in the f-EKF and s-MHE contributes to the applicability (especially the computational efficiency) of the proposed distributed scheme II. We also observe that the proposed scheme gives good performance even with a very small horizon length and its performance is much less sensitive to the horizon size. So it could be much more computationally efficient since a much smaller horizon can be used. If the sampling time decreased for slow subsystem, the number of estimates will increase the accuracy in cost of computational time.

\section*{Conclusions}

In this paper, we developed a distributed state estimation method based on EKF and MHE for a class of two time-scale nonlinear systems, where some of the state variables inhibit both the fast and slow dynamics. The nonlinear system was decomposed into fast and slow subsystems based on singular perturbed parameter $\varepsilon$. In the proposed design, a one-directional communication strategy was established and the method was applied to a chemical process. A series of simulations were carried out to compare the proposed architecture with centralized and decentralised techniques from computational time and accuracy point of view. Owing to different sampling times, the system is a stiff problem where step size plays important role on numerical stability of the solution instead accuracy requirements. However, our design ensures numerical stability, moderate accuracy and low computational time.

\section*{Acknowledgment}

Financial support from Natural Sciences and Engineering Research Council of Canada is gratefully acknowledged.

\renewcommand\refname{Literature Cited}

\newpage~

\newpage~

\newpage~

\newpage~

\newpage~

\newpage~

\begin{center}
\begin{table}[t]
\centering
\caption{Nominal values of parameters of the CSTR}
\begin{tabular}{ |c|c|c| } 
 \hline
$C_{A0}=2.5 \hspace{2mm} \text{mol/l}$ & $c_p=c_{ph}=8.0 \hspace{2mm} \text{J/g K}$ & $\rho=\rho_h=800 \hspace{2mm} \text{g/l}$ \\ \hline

$k_{0}=5 \times 10^{10} \hspace{2mm} s^{-1}$ & $E=60,000 \hspace{2mm} \text{J/mol K}$ & $ \varepsilon=\frac{\rho c_p}{UA}=0.1 \hspace{2mm} \text{s/l} $ \\ \hline

$T_A=305 \hspace{2mm} \text{K}$ & $T_h=330\hspace{2mm} \text{K}$ & $\Delta H_r=20,000 \hspace{2mm} \text{J/mol}$ \\  \hline
$V=1.0 \hspace{2mm} \text{l}$ & $V_h=0.0494 \hspace{2mm} \text{l}$ &\\ \hline
 \end{tabular}
 \label{t1}
 \end{table}
\end{center}

\newpage ~

\begin{center}
\begin{table}[t]
\centering
\caption{Initial states of the process and the initial guesses used in different estimation schemes.
}
\begin{tabular}{ |c|c|c|c|c| } 
 \hline
 $\text{states}$&$\text{Initial condition}$ & $\text{Centralised MHE}$& $\text{f-EKF}$ & $\text{s-MHE}$ \\  \hline
 $C_A \hspace{2mm} (\text{mol/l})$ & $2.5 $ & $1.5$& N/A & $1.5$ \\  \hline
 $C_B \hspace{2mm} (\text{mol/l})$ & $0.0$& $0.0001$& N/A & $0.0001$ \\  \hline
 $T \hspace{2mm} (\text{K})$ & $306$ & $308$& $308$ & $308$ \\  \hline
 $T_j \hspace{2mm} (\text{K})$ & $311$ & $313$& $313$ & $313$ \\  \hline
 \end{tabular}
 \label{t2}
 \end{table}
 \end{center}

 \newpage~

\begin{center}
\begin{table}
\centering
\caption{Estimation performance }
\begin{tabular}{ |c|c|c|c|c| } 
 \hline
 $\sigma_{x_i}$&$\text{Scheme I}$ & $\text{Scheme II}$& $\text{Scheme III}$  \\  \hline
 $\sigma_{C_A}(\%) \hspace{2mm}$ & $12.7$ & $12.7$ & $5.7$   \\  \hline
 $\sigma_{C_B}(\%) \hspace{2mm}$ & $7.15$& $7.15$&  $8.42$ \\  \hline
 $\sigma_T (\%)\hspace{2mm}$ & $0.19$ & $0.55$& $6.43\times 10^{-2}$  \\  \hline
 $\sigma_{T_j}(\%) \hspace{2mm}$ & $2.058\times 10^{-3}$ & $0.667$& $7.192\times 10^{-2}$\\  \hline\hline
 Average RMSE (\%)& $2.88$ & $3.26$& $2.2$\\\hline
 Simulation time (s) & $1.39$ & $1.28$& $5.89$\\\hline 
 \end{tabular}
 \label{t3}
 \end{table}
 \end{center}

\end{document}